\documentclass[sigconf]{acmart}

\usepackage{xcolor} 
\usepackage{multirow} 
\usepackage{fancyvrb} 
\usepackage{subcaption} 
\usepackage{fancyvrb} 

\AtBeginDocument{%
  }

\copyrightyear{2025}
\acmYear{2025}
\setcopyright{rightsretained}
\acmConference[CHI EA '25]{Extended Abstracts of the CHI Conference on Human Factors in Computing Systems}{April 26-May 1, 2025}{Yokohama, Japan}
\acmBooktitle{Extended Abstracts of the CHI Conference on Human Factors in Computing Systems (CHI EA '25), April 26-May 1, 2025, Yokohama, Japan}\acmDOI{10.1145/3706599.3720284}
\acmISBN{979-8-4007-1395-8/2025/04}

\begin{document}

\title{AdaptAI: A Personalized Solution to Sense Your Stress, Fix Your Mess, and Boost Productivity}
\author{Rushiraj Gadhvi}
\authornote{These authors contributed equally to this work.}
\email{rushiraj.gadhvi@plaksha.edu.in}
\orcid{0009-0007-7976-5506}
\affiliation{%
  \institution{Plaksha University}
  \city{Mohali}
  \country{India}
}

\author{Soham Petkar}
\authornotemark[1]
\email{soham.petkar@plaksha.edu.in}
\orcid{0009-0005-9329-5892}
\affiliation{%
  \institution{Plaksha University}
  \city{Mohali}
  \country{India}
}

\author{Priyansh Desai}
\authornotemark[1]
\email{priyansh.desai@plaksha.edu.in}
\orcid{0009-0009-8769-6912}
\affiliation{%
  \institution{Plaksha University}
  \city{Mohali}
  \country{India}
}

\author{Shreyas Ramachandran}
\email{shreyas@plaksha.edu.in}
\orcid{0009-0009-3780-2174}
\affiliation{%
  \institution{Plaksha University}
  \city{Mohali}
  \country{India}
}

\author{Siddharth Siddharth}
\email{siddharth.s@plaksha.edu.in}
\orcid{0000-0002-1001-8218}
\affiliation{%
  \institution{Plaksha University}
  \city{Mohali}
  \country{India}
}
\renewcommand{\shortauthors}{R. Gadhvi, S. Petkar, P. Desai, S. Ramachandran, S. Siddharth}

\begin{abstract}
Personalization is a critical yet often overlooked factor in boosting productivity and well-being in knowledge-intensive workplaces to better address individual preferences. Existing tools typically offer uniform guidance, whether auto-generating email responses or prompting break reminders, without accounting for individual behavioral patterns or stress triggers. We introduce AdaptAI, a multimodal AI solution combining egocentric vision and audio, heart and motion activities, and the agentic workflow of Large Language Models (LLMs) to deliver highly personalized productivity support and context-aware well-being interventions. AdaptAI not only automates peripheral tasks (e.g., drafting succinct document summaries, replying to emails, etc.) but also continuously monitors the user’s unique physiological and situational indicators to dynamically tailor interventions, such as micro-break suggestions or exercise prompts, at the exact point of need. In a preliminary study with 15 participants, AdaptAI demonstrated significant improvements in task throughput and user satisfaction by anticipating user stressors and streamlining daily workflows.
\end{abstract}




\maketitle
\section{INTRODUCTION}
For information workers, maintaining productivity requires a nuanced balance between cognitive demands, physical and emotional well-being, and environmental factors \cite{vinson2024sustainableworkplacementalhealth,patwardhan2016emofitaffectmonitoringsedentary,Graziotin_2013,Constantinides_2020}. While recent advances in intelligent systems---such as conversational agents and context-aware platforms---have shown promise in streamlining tasks \cite{jaber2024cooking,wang2023enabling}, these tools provide only generic or one-size-fits-all recommendations. This is because these artificial intelligence (AI) systems are trained as a whole on digitized human knowledge but are unable to personalize their recommendations for each user's physical and physiological states. Consequently, they often fail to account for user-specific nuances, including personal stressors, work routines, and behavioral tendencies. Past literature discusses this gap by proposing a user-centric approach that utilizes telemetry data gathered via Viva Insights to provide tailored assistance \cite{Nepal_2024}. Yet, such methods do not utilize a multi-modal approach, and thus, do not provide a holistic understanding of users' workplace and their personal needs. This gap hampers the creation of truly personalized productivity solutions.

\begin{figure*}[htp]
  \centering
  \includegraphics[width=\linewidth]{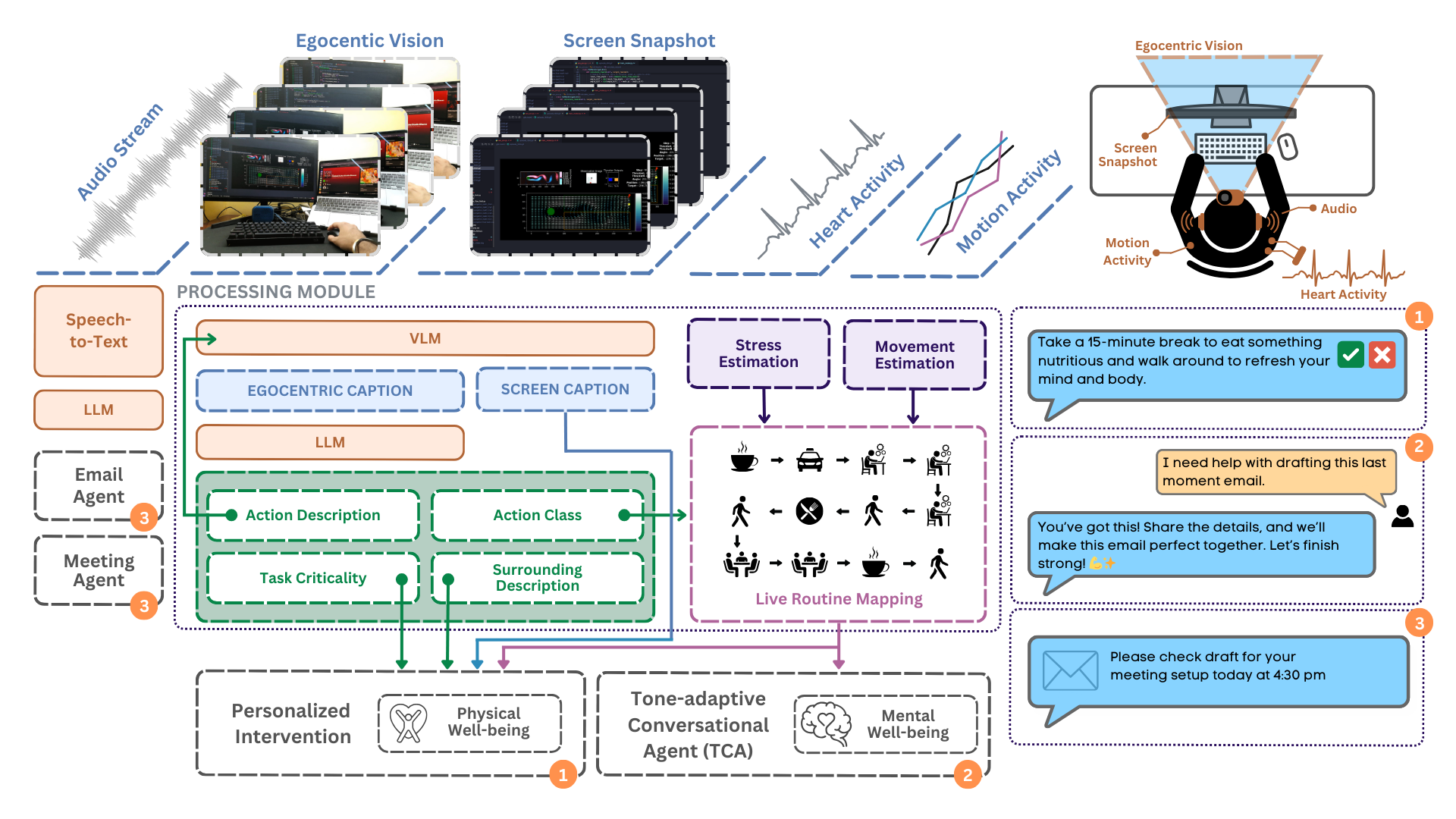}
  \caption{\textbf{AdaptAI's Architecture.} (1) Processing Module integrates real-time streams from vision, audio, motion, and heart activity data, creating LLM-compatible representations; (2) External Task Agents; (3) Personalized Well-being Intervention Pipeline leverages vision insights for impactful physiological interventions; (4) Tone-Adaptive Conversation Agent (TCA) dynamically adjusts tone using heart activity data to address psychological states and support task queries.}
  \Description{A high-level architecture diagram of AdaptAI, showing components including a processing module, external task agents, a personalized well-being intervention pipeline, and a tone-adaptive conversation agent.}
  \label{fig:AdaptAI}
\end{figure*}

To address this challenge, we introduce AdaptAI, a solution leveraging multimodal sensing (e.g., egocentric vision, audio, heart activity, and motion) and harnessing the agentic workflow of Large Language Models (LLMs) to deliver personalized and real-time workplace productivity recommendations and well-being support. AdaptAI integrates three key dimensions of personalization that contemporary AI assistants do not utilize. First, it models the user's physical behavior and the criticality associated with the task being performed to suggest interventions promoting an active work routine and boosting task-specific productivity. Second, it monitors physiological biomarkers---including stress and fatigue indicators derived from vision-based and heart activity-based measurements \cite{alhussein2023wearable}---to deliver personalized interventions. These interventions, ranging from suggesting micro-breaks to recommending brief physical activities, are designed to proactively mitigate stress and foster a sustainable work rhythm boosting workplace well-being. Third, it has the ability to detect multitasking---a characteristic feature of the life of an information worker---and its efficiency in providing personalized recommendations to boost productivity in such a scenario.

Unlike conventional productivity or well-being tools that have previously employed LLM-based static, generic approaches \cite{Oh2021ASR,doi:10.1177/0963721413495872,doi:10.4278/0890-1171-12.1.38,info:doi/10.2196/40789} or the dynamic ones which adhere to just one sensing modality \cite{nepal2024usersurveystelemetrydrivenagents,wu2024ilikesunniei,Nepal_2024,swain2024sensible,wu2024peergpt,liu2024compeergenerativeconversationalagent}, AdaptAI dynamically adapts to individual behavioral and physiological states in real-time using multiple modalities offering a truly personalized user experience.

In a preliminary study (Section ~\ref{sec:study_design}), AdaptAI significantly improved user performance metrics as well as self-reported user satisfaction \cite{grover2020design}. By responding to individual cues and contextual factors, AdaptAI not only minimized productivity bottlenecks but also catered to the growing demand for AI solutions prioritizing user-centric care. Our results underscored AdaptAI's potential to bridge the gap between workplace well-being and personalization, marking a step toward future technologies that serve as personalized partners, rather than merely generic recommendation tools. AdaptAI's codebase has been released as open source. \footnote{https://github.com/gadhvirushiraj/AdaptAI}

\section{RELATED WORKS}
The interplay between users' stress, physical states, and their surrounding environments has been extensively studied for its impact on productivity. Research indicates that elevated stress levels can significantly diminish cognitive performance and overall work efficiency \cite{gjoreski2017continuous, picard2000effective}. Additionally, individuals' physical condition, including factors such as fatigue and health status, directly influences their ability to maintain sustained productivity \cite{meeus2019reduced}. Environmental aspects, such as commuting distance to work, have also been shown to affect employees' productivity and well-being \cite{dam2019towards, taboada2023artificial}. Longer commutes can lead to increased stress and reduced time for rest and personal activities, thereby negatively impacting work performance and job satisfaction \cite{bansal2020optimizing}.

Advancements in bio-sensing technologies have enabled the accurate determination of users' psychological states, including stress and overall well-being. Wearable devices that monitor physiological metrics such as heart rate variability (HRV) and galvanic skin response (GSR) provide reliable indicators of an individual's stress and fatigue levels \cite{gjoreski2017continuous, wei2022chain}. These biosensors facilitate real-time assessments of emotional and cognitive states, offering valuable insights into users' mental health \cite{picard2000effective}. By continuously tracking these biomarkers, systems can detect fluctuations in stress and well-being, enabling timely interventions \cite{meeus2019reduced}.

Before LLMs, various systems were developed to enhance workplace productivity through targeted implementations. These early solutions focused on automating repetitive tasks, optimizing workflow processes, and facilitating team collaboration \cite{dam2019towards, taboada2023artificial, bansal2020optimizing}. For instance, project management tools leveraged AI to streamline task assignments and monitor progress, while decision-making systems provided data-driven recommendations to support managerial choices \cite{jarrahi2018artificial}. Additionally, cognitive workload measurement techniques in Human-Computer Interaction (HCI) employed questionnaires and think-aloud protocols to assess mental demands and adapt user interfaces accordingly \cite{hart2006nasa, brewster1994design}. These pre-LLM systems laid the groundwork for more sophisticated productivity-enhancing technologies by addressing specific aspects of workplace efficiency and employee well-being.

The advent of LLMs has revolutionized the development of AI assistants aimed at improving workplace productivity by leveraging advanced reasoning and contextual understanding capabilities. Recent implementations, such as the Personal Health Insights Agent (PHIA) combine LLMs with wearable data to deliver personalized health insights, demonstrating the potential for interactive and contextually relevant support \cite{merrill2024transforming}. LLMs can facilitate the processing of multimodal data, enabling systems to generate tailored recommendations based on comprehensive behavioral and physiological metrics \cite{minaee2024large, kojima2022large}. These models support chain-of-thought reasoning, which enhance their ability to interpret complex user intents and provide meaningful interventions \cite{brown2020language, wei2022chain}. By incorporating LLMs, contemporary systems surpass previous implementations in their ability to adapt dynamically to individual cognitive workloads and deliver sophisticated, context-aware support for enhancing workplace performance.

However, to the best of our knowledge, existing workplace productivity and well-being enhancement systems do not 1) provide a holistic overview of an information worker's workplace from their egocentric view as well as their surroundings, 2) assess the criticality of tasks being undertaken by the user to accordingly tailor and prioritize the suggested interventions, and 3) deliver personalized user feedback based on their physical and physiological well-being. These can only be done by taking a multi-modal approach combining the power of LLMs with egocentric audio-visual systems (to understand if the user is multitasking and the tasks that the user may be engaged in), motion sensors (to assess the user's physical activity), and bio-sensing modalities (to model a user's physiology such as by measuring heart and brain activity). Previous works \cite{nepal2024usersurveystelemetrydrivenagents,wu2024ilikesunniei,Nepal_2024,swain2024sensible,wu2024peergpt,liu2024compeergenerativeconversationalagent}
have attempted to provide methods addressing either adaptations or interventions through conversational interfaces, dashboards, and similar mechanisms. Nevertheless, none of these systems have combined bio-sensors and egocentric vision, which we consider essential for capturing the extensive range of data required for effective personalization. The integration of these diverse data sources is crucial for understanding the multifaceted aspects of user behavior and environmental interactions, thereby enabling more comprehensive and tailored interventions to enhance workplace productivity and well-being.

\section{ADAPTAI SOLUTION}

AdaptAI Architecture (Figure \ref{fig:AdaptAI}) comprises of four main components: (1) Processing Module, which has a sensing layer capturing real-time streams from three different modalities: vision (egocentric camera and screen snapshot), microphone, and an electrocardiogram (ECG) strap band or a smartwatch capable of providing heart activity data. This module is also responsible to create information representation that can be easily reasoned upon using an LLM. More modalities enable the system to access the user's current state more holistically (2) External Task Agents, automating simple extra tasks (3) Personalized Well-being Intervention Pipeline, utilizing insights extracted from vision modality to generate environmentally feasible and impactful interventions to support the physiological state of the user (4) Tone-adaptive Conversation Agent (TCA) addressing the psychological state of the user, utilizing heart activity analysis to adapt tone of the conversational agent dynamically, incorporating motivational elements during stress periods, while providing solutions to user's task queries. 

\begin{figure*}[t!]
  \centering
  \includegraphics[width=\linewidth]{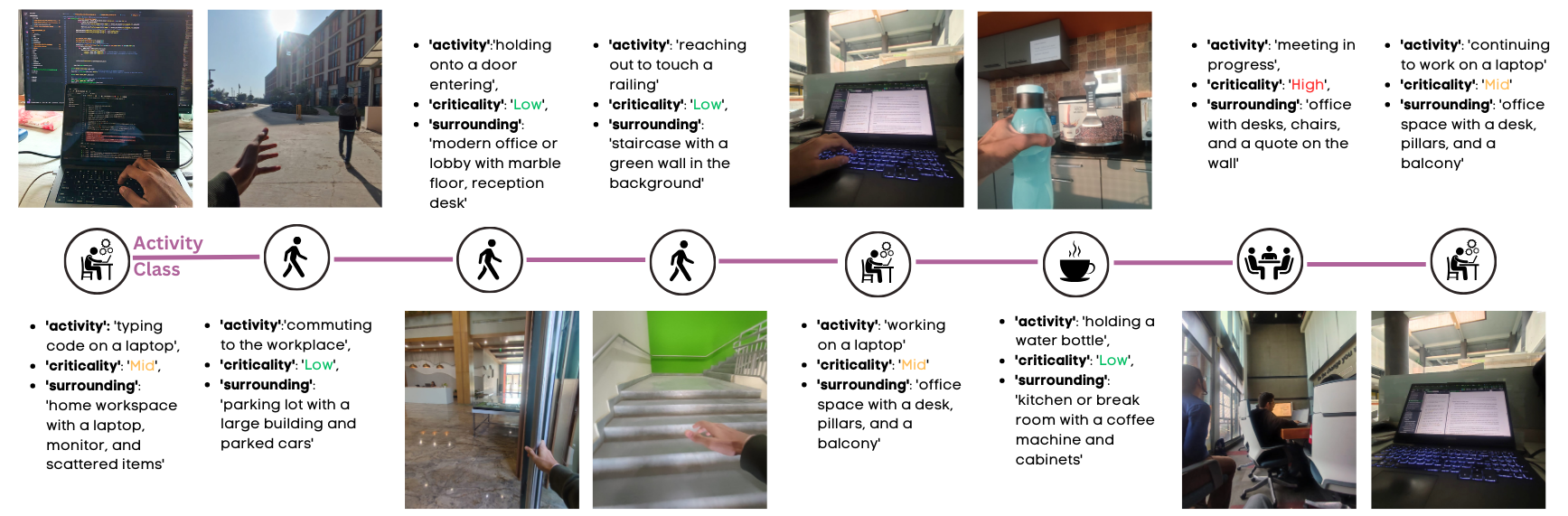}
  \caption{\textbf{Example of a User's Daily Routine.} The figure illustrates AdaptAI's detection of activity descriptions, classes, surrounding context, and criticality.}
  \Description{The image illustrates AdaptAI's detection of user activities throughout the day, showing various tasks like working, commuting, and meetings, along with their context and criticality levels.}
  \label{fig:AdaptAI-rtable}
\end{figure*}

\subsection{Processing Module}

\subsubsection{Vision-based Pipeline}
Vision is used to gather essential information about the user’s activities, their workplace surrounding, and on-screen task. Using an egocentric camera, we capture frames at \textit{5-second} intervals, resizing them to 640×480 pixels before sending them to a vision-language model (VLM)\cite{bordes2024introduction}. To describe what is happening in each frame, we used the open-source \textit{llama-3.2-11b-vision} model to generate detailed captions. While the VLM is highly capable of extracting insights directly from the frames, we noticed some inconsistencies during testing. To address this, we pass the generated captions to an open-source LLM \textit{llama-3.1-8b}, which then extracts various insights from the captions as shown in Figure \ref{fig:AdaptAI-rtable}. The following four insights are extracted :

\begin{itemize}
\item \textbf{Activity Description}: We aim to capture the specific actions the user is performing in detail, including their interactions with objects or devices, such as typing, moving items, or gestures. If the user is interacting with a laptop, it also tries to identify the tasks being performed on the screen, such as browsing, coding, or video conferencing.

\item \textbf{Activity Class}: We also categorize the user's activities into more coarse tasks, such as whether they are doing desk work, commuting, eating, in-meeting, or socializing. This broader understanding allows the system to distinguish between different states of activity, as creating an accurate routine is crucial to generating accurate interventions.

\item \textbf{Task Criticality}: Assesses the criticality of the user’s current activity to adjust interactions and prioritize notifications. \textcolor[rgb]{0,0.59,0}{Low} criticality involves routine tasks requiring minimal focus, like washing hands or drinking water. \textcolor[rgb]{1,0.65,0}{Mid} criticality pertains to activities needing moderate attention, such as walking in varied environments or casual conversations. \textcolor[rgb]{0.86,0.13,0.13}{High} criticality includes tasks demanding intense focus, like presenting, in meetings, complex work, or handling emergencies.

\item \textbf{Surrounding Description}: Refers to the broader context of the user's environment, including factors like location, the presence of other people, and the overall setting. It identifies whether the user is in a workspace, home, or public place, accounting for potential distractions.
\end{itemize}

The VLM excels at detecting objects and inferring user activities but sometimes misinterprets objects. For instance, detecting objects \textit{`hand-dryer'} as a \textit{`food-dispenser'} affects the extracted descriptions. To resolve such errors in caption description, we pass past detected \textit{fine-activity} along with the frame, i.e., for the earlier example we pass activity of \textit{`washing hands in a basin'} along with the frame, as shown in Figure \ref{fig:correct_caption_samples} (Appendix). This approach has proven highly effective in improving performance. In addition to processing egocentric frames, the system also captures snapshots of the user’s primary device screen (if available). The pre-processing methodology and models used for these snapshots are consistent with those employed for egocentric frames. The VLM extracts insights about on-screen activities, like browsing, coding, or creating presentations, to build multitask based interventions downstream. Descriptions of screen activities from the last ten snapshots are stored and updated in sync with the egocentric camera's insight extraction interval. Appendix \ref{sec: prompts_vision} provides the prompts used in the vision-based pipeline.

\subsubsection{Stress and Movement Estimation}
Maintaining low-stress levels in the workplace is important for ensuring productivity and overall well-being. The AdaptAI solution incorporates a stress estimation module that utilizes data from the heart activity captured either through a chest positioned ECG sensor or a smartwatch capable of providing it. The ECG data such obtained is recorded in millivolts at a sampling frequency of 200 Hz. To assess the user's stress state, HRV features are computed, specifically the pNN50 metric, which represents the percentage of successive inter-beat (RR) intervals differing by more than 50 milliseconds. Threshold values are defined to categorize stress levels: if the user's pNN50 is below 20, the user is classified as being under \textit{high stress}; if the pNN50 is between 20 and 50, the user is considered to be under \textit{moderate stress}; and if the pNN50 is above 50, the user is classified as experiencing \textit{low stress} \cite{HRV_Stress}. In addition to HRV analysis, the mean heart rate is calculated for the same time interval to provide supplementary insights into the user's physiological state. For movement estimation, the inertial measurement unit (IMU) is used to calculate the number of steps taken by the user during predefined time intervals. The step count serves as a proxy for assessing walking activity and commuting within the workplace, offering valuable insights into the user’s physical activity patterns.

\subsubsection{Routine Table Generation}
Through iterative experimentation, we identified routine generation as the most optimal and concise method for delivering information to LLMs. This approach provides sufficient granularity for lightweight LLMs to process efficiently, without compromising output quality. Insights extracted from the vision pipeline and stress estimation are stored in a temporary database, which is processed at regular intervals of \textit{15 minutes} to generate a Routine Table. An example of the Routine Table is provided in Table \ref{tab:routine_table} (Appendix). The Routine Table records the duration (in minutes) that users spend on each course activity class within the specified time intervals. While we experimented with intervals of \textit{30-minute}, \textit{1-hour}, we found \textit{15 minutes} interval to be effective in achieving a balance for information granularity. The intervals follow a 24-hour format, enabling downstream LLMs to infer whether specific routines, such as lunch or dinner, have been completed or remain pending.

\subsection{External Task Agents}
AdaptAI features an audio processing pipeline to direct user attention to core tasks by handling minor tasks via an agent. Audio streams divided into one-minute segments are processed by the \textit{OpenAI Whisper} model, specifically \textit{whisper-large-v3-turbo}, for speech-to-text conversion. The resulting text is then scrutinized for actionable items. The system incorporates an agentic strategy to automate standard tasks informed by the transcribed text, activating tools like email generation or calendar scheduling. This also allows users to attach their own agents in the future to provide customized automation as per their needs.

\subsection{Personalized Intervention Pipeline}
\begin{figure}[htp]
  \includegraphics[width=0.5\textwidth]{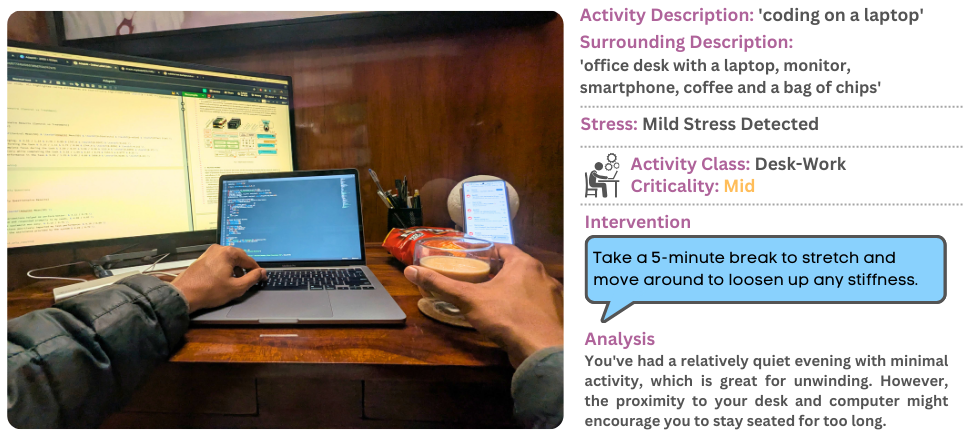}
  \caption{\textbf{AdaptAI displays real-time Intervention, while correctly accessing the conditions and extracting insights.}  
  }
  \Description{Demonstration image illustrating AdaptAI detecting a user coding at a desk, assessing mild stress, and suggesting a 5-minute break to prevent stiffness.}
  \label{fig:Intervent-AdaptAI}
\end{figure}

The primary objective of the Personalized Intervention Pipeline is to improve the user's physiological state by delivering timely and appropriate interventions (Figure \ref{fig:Intervent-AdaptAI}). For instance, a user doing coding for multiple hours is intervened to take a small 5-minute break, stretch, and walk around. It leverages the Routine Table as its foundational data source, utilizing all elements of the Routine Table except HRV metrics, it also integrates instantaneous contextual information of screen activities from the vision pipeline to enhance awareness of tasks done by users. Experiments indicated that the smaller \textit{llama-3.1-8b} model exhibited inconsistent performance, motivating an upgrade to the more robust and bigger open-source \textit{llama-3.1-70b} model. Since interventions are not made frequently, this upgrade to a larger model does not affect the intervention time. By leveraging the capabilities of the language model \textit{llama-3.1-70b}, the pipeline's prompt (Appendix \ref{sec: prompts_intervention}) was designed to generate interventions that are both creative and mindful. The creative aspect ensures interventions adapt dynamically to the user's surroundings and conditions, promoting physical well-being. Mindfulness ensures interventions are accurate, feasible, and contextually appropriate. Furthermore, the pipeline provides an analysis along with interventions allowing users to understand the LLM's reasoning behind each intervention made and decide whether to accept or reject it based on this rationale. The pipeline operates synchronously with the Routine Table, processing data and generating interventions at intervals of \textit{15 minutes}. To avoid disruption during high-criticality situations, such as meetings, conversations, or commuting, the pipeline incorporates criticality assessment from the vision pipeline. Interventions are permitted only when the task criticality is assessed as 
\textcolor[rgb]{0,0.59,0}{Low} or \textcolor[rgb]{1,0.65,0}{Mid}. In cases of \textcolor[rgb]{0.86,0.13,0.13}{High} criticality (ongoing meeting/presentation), interventions are blocked until the criticality level decreases, ensuring user focus and minimizing unnecessary interruptions.

\subsection{Tone-adaptive Conversation Agent (TCA)}
The Tone-adaptive Conversational Agent (TCA) is designed to address the psychological state of the user by dynamically adjusting its tone based on stress levels. The TCA employs the \textit{llama-3.1-70b} model to generate responses to the user queries while providing it in a mental well-being-supporting tone. The TCA achieves a high degree of personalization by leveraging the user’s Routine Table, which provides detailed insights into the user’s physical and mental state, derived from activity tracking and physiological data. TCA is accessible to the user at any time through a dedicated user interface (Figure \ref{fig:ui_TCA} in Appendix).


The base prompt (Appendix \ref{sec: prompts_TCA}) incorporates a two-part tone adaptation strategy, designed to dynamically adjust to the user's current state and the progression of the conversation. These two components are detailed below:
\begin{itemize}
\item \textit{Dynamic Tone Adjustment:} When a user engages with the conversational agent, their current stress level determines the tone of the response. If the stress level is high, the agent adopts a motivating and encouraging tone. Conversely, during periods of low stress, the tone becomes neutral and straightforward. This ensures that the communication aligns with the user’s psychological needs, promoting a sense of comfort and support.
\item \textit{Progressive Simplification of Tone:} As the conversation progresses, the tone gradually shifts toward a normal and more direct style. This transition aims to minimize the cognitive load on the user, facilitating clarity and ease of understanding, particularly during extended interactions.
\end{itemize}

For example, if a user has spent a long workday debugging code and exhibits high stress, the TCA will respond to their queries with motivational and positive language. This personalized adaptation is informed by the user's physical state, stress levels, and recent activities recorded in the Routine Table, allowing the TCA to recognize potential fatigue and tailor its responses accordingly (Figure \ref{fig:ui_TCA} in Appendix). While the agent can handle a wide range of tasks similar to a general-purpose conversational AI system like ChatGPT, its key distinction is this ability to personalize responses in real-time based on the user's state.

\begin{table*}[t!]
\centering
\caption{Comparison of Control and AdaptAI Groups on Likert Scale Responses ranging from 1 (Strongly Disagree) to 5 (Strongly Agree). Statistical analysis includes the U-statistic, p-value from the Mann-Whitney U test, and effect size (Cohen’s d).}
\resizebox{\linewidth}{!}{
\begin{tabular}{l c c c c c}
\toprule
\textbf{Question} & \textbf{Control Mean/SD} & \textbf{AdaptAI Mean/SD} & \textbf{U-Statistic} & \textbf{p-value} & \textbf{Effect Size} \\
\midrule
I found this to be challenging. & 3.25 / 1.14 & 2.78 / 0.89 & 1747.0 & \textbf{0.0247} & \textbf{0.45} \\
I felt stressed while performing the task. & 3.16 / 1.11 & 2.72 / 0.86 & 1744.0 & \textbf{0.0256} & \textbf{0.45} \\
I was able to maintain complete focus during the task. & 3.59 / 0.97 & 3.98 / 0.58 & 1122.0 & \textbf{0.0399} & \textbf{-0.49} \\
I managed my time effectively while completing the task. & 3.33 / 1.04 & 3.67 / 0.76 & 1152.5 & 0.0777 & -0.38 \\
I am satisfied with my performance in the task. & 3.28 / 1.25 & 3.85 / 0.89 & 1058.0 & \textbf{0.0230} & \textbf{-0.52} \\
\bottomrule
\end{tabular}
}
\label{tab:post_task_results}
\end{table*}

\begin{table*}[!ht]
\centering
\caption{Additional Questionnaire Responses from the AdaptAI Group: The table highlights the mean and standard deviation (Mean/SD) of responses exclusively from the treatment group, evaluating their perceptions of personalized interventions, system responsiveness, ease of interaction, impact of interventions, and satisfaction with the assistance provided by the system.}
\begin{tabular}{lcc}
\toprule
\textbf{Question} & \textbf{AdaptAI Mean/SD} \\
\midrule
The personalized interventions helped me perform better. & 4.12 / 0.78 \\
The system understood and responded promptly to my needs. & 4.08 / 0.82 \\
Interacting with the system/TCA was easy. & 4.15 / 0.75 \\
The system interventions positively impacted my task performance. & 4.10 / 0.80 \\
I am satisfied with the assistance provided by the system & 4.20 / 0.72 \\
\bottomrule
\end{tabular}
\label{tab:treatment_only_results}
\end{table*}

\section{STUDY DESIGN AND EVALUATION}
\label{sec:study_design}
\subsection{Procedure}
\begin{figure*}[!t]
    \centering
    \begin{subfigure}[b]{0.42\textwidth}
        \centering
        \includegraphics[width=\textwidth]{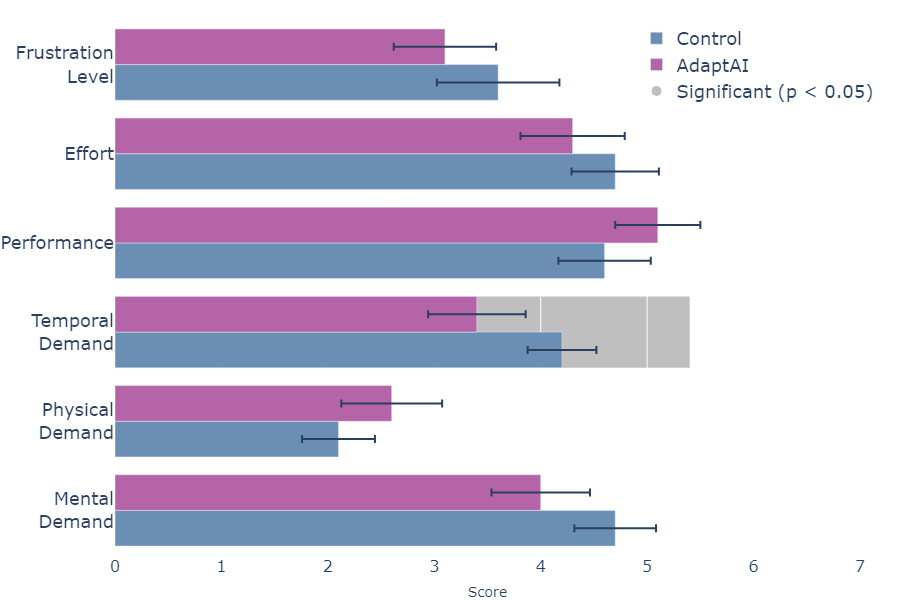}
        \caption{Math Test}
        \Description{Graph showing NASA-TLX ratings for the Math Test.}
        \label{fig:graph1}
    \end{subfigure}
    \hfill
    \begin{subfigure}[b]{0.42\textwidth}
        \centering
        \includegraphics[width=\textwidth]{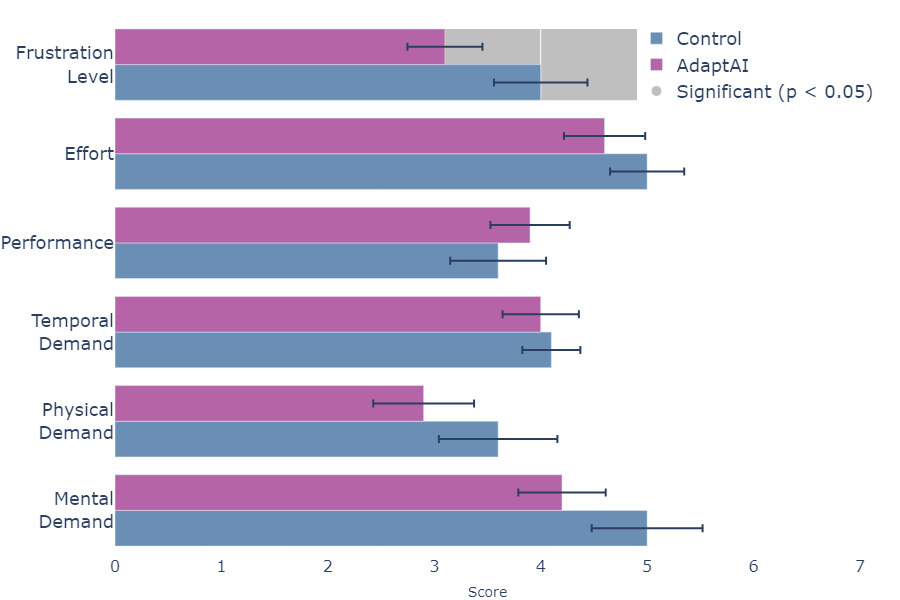}
        \caption{Webpage Coding Test}
        \Description{Graph showing NASA-TLX ratings for the Webpage Coding Test.}
        \label{fig:graph2}
    \end{subfigure}

    \begin{subfigure}[b]{0.42\textwidth}
        \centering
        \includegraphics[width=\textwidth]{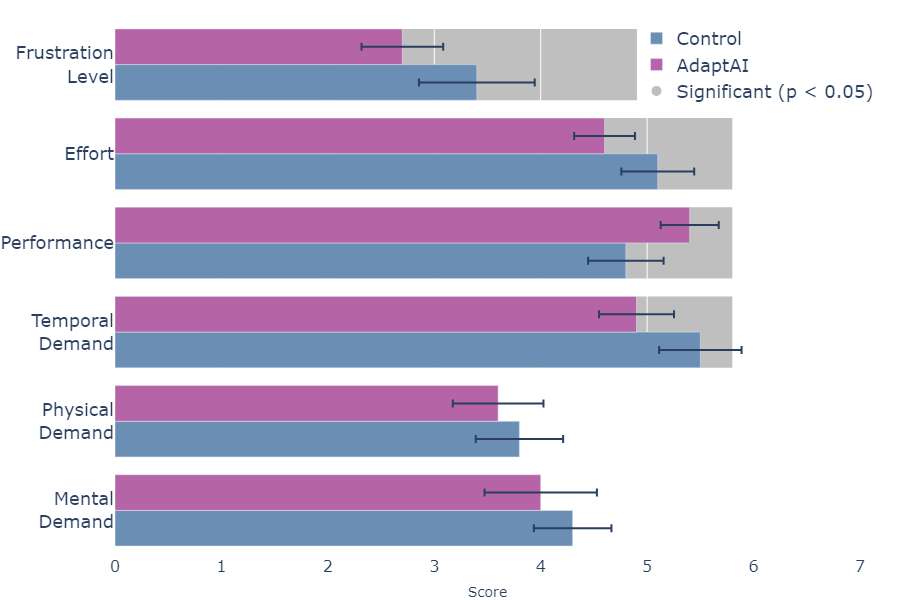}
        \caption{Typing Test}
        \Description{Graph showing NASA-TLX ratings for the Typing Test.}
        \label{fig:graph3}
    \end{subfigure}
    \hfill
    \begin{subfigure}[b]{0.42\textwidth}
        \centering
        \includegraphics[width=\textwidth]{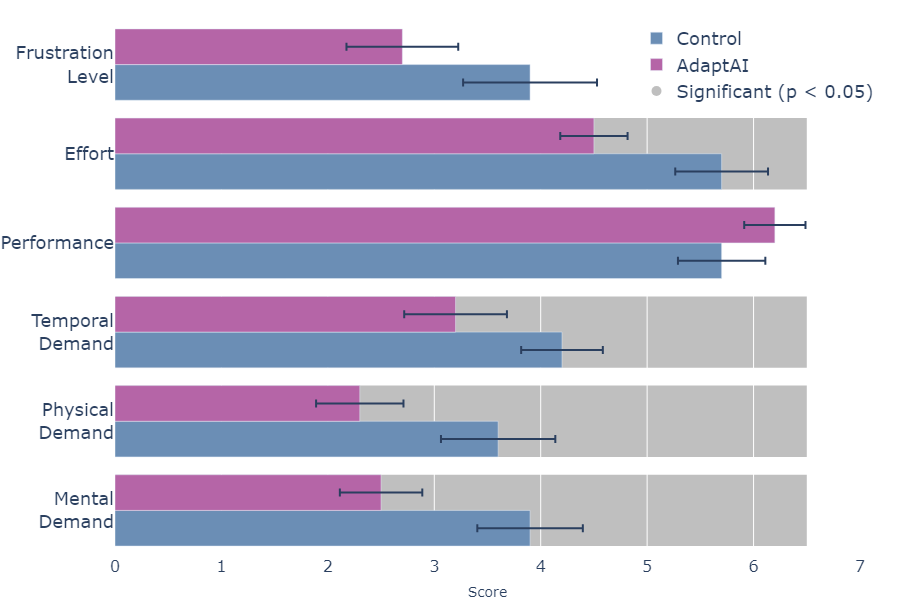}
        \caption{Data Entry Test}
        \Description{Graph showing NASA-TLX ratings for the Data Entry Test.}
        \label{fig:graph4}
    \end{subfigure}
    \caption{\textbf{Participants’ ratings on NASA-TLX questions
(scale: 1-low to 7-high) for the 4 tasks
in control and AdaptAI groups. All highlighted (in gray) are rating differences that are statistically significant with \textit{p < 0.05} via Wilcoxon
signed-rank tests.}
}
    \label{fig:results}
\end{figure*}

This research evaluated AdaptAI's real-time effects on stress reduction and task performance improvement across diverse tasks. Fifteen professionals (10 males, 5 females), aged 19 to 42 (M = 25.13, SD = 7.48), participated, selected for their relevant work experience and stress encounters \cite{pascoe2020impact, robotham2006stress}. Using a within-subjects design, the study had Control and Treatment phases. In the Control phase, participants employed \textit{llama-3.1-70b} conversation agent (CA) without AdaptAI. In the Treatment phase, they used AdaptAI, featuring TCA and interventions. Participants undertook four tasks: a math test, a typing test, webpage coding (webpage design using HTML/CSS) for e-commerce, and data entry. The math task took approximately 10 minutes, the typing task took 5 minutes, the design task took 15 minutes, and the data entry task took 10 minutes. During Control, tasks were completed unaided, followed by filling out the NASA Task Load Index (NASA-TLX), a subjective workload assessment tool measuring mental, physical, and temporal demands, performance, effort, and frustration levels. Additionally, participants completed a Likert Scale Index questionnaire, a psychometric scale used to assess subjective perceptions of workload, stress, and performance, as shown in Figure ~\ref{fig:results}, Table~\ref{tab:post_task_results}, and Table~\ref{tab:treatment_only_results}.


During the Treatment phase, participants completed identical tasks while engaging with AdaptAI, which offered task-specific advice to minimize stress and improve performance. They could also interact with the TCA for added assistance. Following each task in this phase, participants completed the NASA-TLX and Likert Scale questionnaires to assess the interventions' effectiveness and the overall utility of the personalized support from AdaptAI. The study was conducted in a controlled, standardized setting to maintain consistency. Participants were seated in a well-lit room, simulating an office environment, to ensure comfort and focus during tasks. Interventions were presented as on-screen prompts within a text editor, allowing participants to comprehend the suggested actions. Data collected were analyzed to contrast control and intervention conditions, offering a thorough quantitative assessment of AdaptAI’s role in reducing stress and enhancing performance. The NASA-TLX questionnaire recorded responses on mental, physical, and temporal demands, along with performance, effort, and frustration (Figure ~\ref{fig:results}). For both Control and Treatment conditions, the Likert Scale measured factors like task challenge, stress, focus, time management, and satisfaction (Table \ref{tab:post_task_results}). In the Treatment phase, extra items on the Likert Scale evaluated AdaptAI’s support, promptness, and effectiveness (Table~\ref{tab:treatment_only_results}).

\subsection{Results}
NASA-TLX workload ratings across four tasks revealed AdaptAI's effectiveness in reducing workload and enhancing user experience. Significant reductions were observed in \textbf{Mental Demand} (\textit{p} = 0.011), \textbf{Physical Demand} (\textit{p} = 0.005), \textbf{Temporal Demand} (\textit{p} = 0.031), and \textbf{Effort} (\textit{p} = 0.037) during the Data Entry task, highlighting reduced cognitive and physical strain.

In the Typing Test, AdaptAI significantly reduced \textbf{Temporal Demand} (\textit{p} = 0.014), \textbf{Performance} (\textit{p} = 0.014), \textbf{Effort} (\textit{p} = 0.025), and \textbf{Frustration Level} (\textit{p} = 0.035). These results highlight AdaptAI’s ability to alleviate time pressure, enhance users’ confidence in their performance, reduce physical and mental effort, and relieve emotional stress while performing motor-intensive tasks.

In the Math Test, \textbf{Temporal Demand} (\textit{p} = 0.033) was significantly reduced, demonstrating AdaptAI’s effectiveness in mitigating time pressure in tasks involving numerical reasoning.

In the Webpage Coding task, \textbf{Frustration Level} (\textit{p} = 0.021) was significantly lower, reflecting AdaptAI’s potential to reduce emotional burden in such tasks, while other workload dimensions showed no significant improvement.

Notably, across tasks, \textbf{Temporal Demand} consistently showed improvement, with significant reductions in the Data Entry, Typing Test (\textit{p} = 0.014), and Math Test (\textit{p} = 0.033). These findings collectively demonstrate that the interventions provided by AdaptAI not only reduce emotional and physical stress but also significantly enhance efficiency and confidence in performing cognitively and physically demanding tasks, further supporting its effectiveness as a workload reduction tool.

The post-task Likert scale analysis revealed significant improvements in user focus (Table ~\ref{tab:post_task_results}, satisfaction, and stress reduction when using AdaptAI compared to the control system. Participants reported being able to maintain better focus during tasks with AdaptAI (Control: 3.59 ± 0.97; AdaptAI: 3.98 ± 0.58; U = 1122.0, p = 0.0399, r = -0.49), highlighting its positive impact on attention and engagement. Similarly, satisfaction with task performance significantly improved in the AdaptAI condition (Control: 3.28 ± 1.25; AdaptAI: 3.85 ± 0.89; U = 1058.0, p = 0.0230, r = -0.52). Furthermore, participants experienced reduced stress while performing tasks with AdaptAI (Control: 3.16 ± 1.11; AdaptAI: 2.72 ± 0.86; U = 1744.0, p = 0.0256, r = 0.45), reflecting its effectiveness in alleviating emotional strain. These findings collectively demonstrate that AdaptAI interventions helped users stay focused, reduced stress, and improved satisfaction, ultimately leading to a more positive task experience.

Participants' treatment-only Likert scale responses further emphasize the success of AdaptAI in delivering personalized and accessible interventions (Table ~\ref{tab:treatment_only_results}). High ratings were observed across key questions, with participants agreeing that the system was easy to interact with (4.15 ± 0.75) and understood their needs promptly (4.08 ± 0.82). Moreover, participants felt that the personalized interventions were effective in improving their performance (4.12 ± 0.78) and positively impacted their task execution (4.10 ± 0.80). Overall satisfaction with the system was exceptionally high (4.20 ± 0.72), underscoring AdaptAI’s role as a reliable and user-friendly tool. These results reinforce the notion that a personalized, adaptive system like AdaptAI significantly enhances the user experience, fostering improved task outcomes and user satisfaction.

Participants who gave subjective feedback after the experiment also consistently highlighted AdaptAI’s flexibility in catering to the demands of different tasks, its fast inference speed (see Appendix \ref{sec:Inference_speed}), and its relevance in providing actionable interventions. For creative and cognitively demanding tasks like webpage coding, participants appreciated how AdaptAI encouraged brief moments of reflection and movement, with one participant noting, \textit{``It reminded me to step back and change my perspective, which really helped me solve a design challenge.''} Similarly, during repetitive or motor-intensive tasks like typing, participants found AdaptAI’s suggestions practical and impactful. A participant shared, \textit{``The prompts to relax my shoulders and adjust my posture came just at the right time to relieve tension.''} Notably, for lengthy and monotonous tasks, participants found AdaptAI’s interventions particularly beneficial, as one user remarked, \textit{``It helped me stay refreshed and focused during a long, draining task, which I wouldn’t have managed as effectively on my own.''} This feedback underscores AdaptAI’s ability to adapt dynamically to the nature of the task, making it a valuable tool for improving productivity and well-being across a range of scenarios.

\section{ETHICAL CONSIDERATIONS}
This research places a strong emphasis on upholding ethical standards to protect the rights and welfare of participants. Ethical compliance was maintained throughout the study lifecycle, aligning with best practices in psychological and human-computer interaction research. Volunteers provided explicit written informed consent prior to beginning the study and maintained the freedom to withdraw whenever they chose. All data were fully anonymized at the source, ensuring no personally identifiable information was collected. Anonymization with unique IDs, secure storage, and limited access measures were employed to safeguard data confidentiality. The experiment protocol was reviewed and verified by the institute's ethics committee.


\section{CONCLUSION AND FUTURE WORK}


By introducing AdaptAI, a system that integrates egocentric vision, physiological signals, and agentic workflows powered by LLMs, we demonstrated how personalized interventions can reduce cognitive and physical strain, reduce stress, and improve workplace productivity. Across four tasks, AdaptAI consistently reduced workload dimensions such as Temporal Demand, Effort, and Frustration Level, while enhancing User Satisfaction and Performance. These findings underscore AdaptAI's potential to dynamically adapt to individual needs, paving the way for transformative applications in personalized productivity support and well-being management.

While AdaptAI demonstrates significant potential in reducing workload, certain limitations warrant discussion. In dynamic, multi-person environments, the system's reliance on vision and audio capture raises privacy concerns, particularly when individuals may be recorded without explicit notice. Future work could explore on-device processing and privacy-preserving methods to mitigate privacy risks. The system's reliance on black-box models necessitates the implementation of techniques to ensure more responsible and transparent interventions in future iterations.

This study's evaluation remains limited to short durations (up to one hour) and specific tasks, and does not address long-term performance over several hours. Future research could include longitudinal studies to assess AdaptAI’s sustained impact on workload reduction and productivity enhancement in diverse, real-world settings rather than a within-subjects experiment design . Lastly, the current task set primarily focuses on logical and repetitive tasks such as coding, math, and data entry. To better evaluate AdaptAI’s adaptability across different cognitive workloads, future experiments could incorporate creative tasks as well as collaborative activities in naturalistic work environments.


\bibliographystyle{ACM-Reference-Format}
\bibliography{sample-base}

\appendix

\section{THE PROMPTS IN ADAPTAI}
Please note prompts in this section, have been restructured for better readability.

\subsection{The prompts of Vision-based Pipeline}
\label{sec: prompts_vision}

\subsubsection{Egocentric Captioning Prompt}
 You are a highly trained specialist at captioning egocentric images. You are tasked to generate a detailed description of a scene from an egocentric perspective while adhering to strict observational guidelines.
\begin{itemize}
\item \textbf{Direct Observation}: Describe only what is directly visible from the given perspective. Include visible actions and details of the surroundings without speculating beyond the provided visual context.
\item \textbf{Environmental Details}: Note specific aspects such of env: Lighting conditions (e.g., bright, dim, artificial, natural light). Time of day if discernible (e.g., morning, afternoon, evening). Location context (e.g., indoors, outdoors, room type, or setting).
\item \textbf{Restrictions}: Avoid assumptions or inferences about unseen objects, actions, or circumstances. Do not include subjective opinions, personal feelings, or unverifiable details. If body parts (e.g., hands) are not visible, do not speculate on their position or activity.
\end{itemize}
\textbf{Previous frame detected activity:} "\texttt{\{pre\_frame\_act\}"} \\
\textbf{Example Description}:
"The scene shows a brightly lit office space with overhead fluorescent lighting. A desk is visible with a laptop, a coffee mug, and some scattered papers. The background features a large window evident of natural lightning, indicating it might be late afternoon. No body parts are visible in the frame. "

\subsubsection{Screen Snapshot Captioning Prompt}
You are tasked with analyzing a detailed description of an screen snapshot captured, extract a concise and accurate description of the activity being performed in the scene.

\textbf{Activity Description}: Identify the specific activity visible in the image. Be precise, especially if the activity involves using a laptop or computer, by guessing the exact task (e.g., writing a report, making presentation, solving a Sudoku puzzle, debugging code). The description should capture the essence of the activity clearly and succinctly.
<\textbf{Output Structure:} ... >

\subsubsection{LLM Insight Extraction Prompt} \label{sec:LLM_Insights_Prompts}

Analyze egocentric image descriptions and extract actionable insights across three key dimensions: activity, criticality, and surrounding context.
\begin{itemize}
\item \textbf{Activity}: Identify the action the person appears to be performing in the image. Provide clear, concise descriptions.
\item \textbf{Best Suited Activity Classification}: Choose one from "Desk\_Work" (any work-related), "Commuting" (walking), "Eating" (having lunch, coffee break), "In\_Meeting" (socializing, physical meeting, presentations), "Other"
\item \textbf{Criticality}: Assign a criticality level based on the following definitions: \textbf{Low}: Routine or minimal focus tasks (e.g., drinking water, organizing papers). \textbf{Mid}: Tasks requiring moderate focus (e.g., walking in a crowded space, typing). \textbf{High}: Demanding tasks requiring significant focus (e.g., driving, presenting to an audience).
\item \textbf{Surrounding}: Describe the visible environment or context. Include notable objects or features relevant to understanding the scene.
\end{itemize}
<\textbf{Output Structure:} ... > 

\subsubsection{LLM Insight Extraction Few Shot Example}
\begin{Verbatim}[frame=single]
Description:
The person is sitting at a desk with a laptop open, 
typing on the keyboard. A cup of coffee is 
nearby, and there are papers scattered around.
Output:
[typing on a laptop | Desk_Work | Mid | 
office desk with papers and a coffee cup]
\end{Verbatim}

\begin{Verbatim}[frame=single]
Description:
The person is walking through a corridor while 
looking down at their phone. The hallway is well-lit 
with doors on either side.
Output:
[walking while using a phone | Commuting | Mid | 
office hallway with doors on both sides]
\end{Verbatim}


\subsection{The prompts of Personalized Intervention Pipeline}
\label{sec: prompts_intervention}

\subsubsection{Intervention Prompt}
You are a workplace wellness assistant responsible for evaluating the following inputs and delivering well-structured recommendations that help the user improve his/her physiological state helping them in there task performance and productivity.
\begin{itemize}
    \item \textbf{Stress Level:} Determine the individual's current stress status (\texttt{stressed} or \texttt{not stressed}).
    \item \textbf{Activity Timetable:} Evaluate the hourly distribution of the individual's activities, which include:
    \begin{itemize}
        \item \texttt{time}: Time range (e.g., \texttt{8-9 PM}).
        \item \texttt{desk\_work}: Minutes spent on desk work.
        \item \texttt{commuting}: Minutes spent commuting.
        \item \texttt{eating}: Minutes spent eating.
        \item \texttt{in\_meeting}: Minutes spent in meetings or discussions.
    \end{itemize}
    \item \textbf{Surrounding Type:} Identify the individual’s current environment (e.g., \texttt{cubicle}, \texttt{meeting room}, \texttt{office}).
    \item \textbf{Screen Capture Data:} Summarize the user's real-time activities based on screen observations (e.g., \texttt{"Working on a spreadsheet"}, \texttt{"Debugging Python code"}). These provide insights into multitasking tendencies and workflow patterns.
\end{itemize}

\textbf{Output Structure:}
\begin{itemize}
    \item \textbf{Analysis:} A comprehensive evaluation of the individual's state, identifying patterns such as frequent context-switching, prolonged inactivity, or inefficient task prioritization.
    \item \textbf{Interventions:} Quick interventions considering the user's current surroundings and physical state.
\end{itemize}

\subsubsection{Few Shot Examples of Personalized Intervention Pipeline}
\begin{itemize}
\item \textbf{Example 1} \\ \begin{Verbatim}[frame=single]
Input:{
    "stress_level": "stressed",
    "activity_timetable": """
        Time,Desk Work (min),Commuting (min),
        Eating (min),In-Meeting (min)
        8-9 AM,60,0,0,0
        9-10 AM,45,0,15,0
        10-11 AM,40,0,0,20
    """,
    "surrounding_type": "cubicle",
    "screen_capture_data": """
        Frame 1: Working on a spreadsheet.
        Frame 2: Browsing emails.
        Frame 3: Debugging Python code.
    """
}
Output:{
  "Analysis": "Frequent context-switching
  between different tasks leads to reduced 
  focus and increased stress.",
  "Task Improvement": "Allocate dedicated time
  blocks for focused work to minimize 
  distractions.",
  "Interventions": {
    "Immediate Action": "Take a 5-minute break 
    with deep breathing exercises.",
    "Follow-Up": 
    "Implement the Pomodoro technique to
    balance work and rest intervals."
  }
}
\end{Verbatim}

\item \textbf{Example 2} \\ \begin{Verbatim}[frame=single]
Input: {
    "stress_level": "not stressed",
    "activity_timetable": """
        Time,Desk Work (min),Commuting (min),
        Eating (min),In-Meeting (min)
        7-8 AM,20,30,10,0
        8-9 AM,50,0,10,0
        9-10 AM,40,0,0,20
    """,
    "surrounding_type": "office",
    "screen_capture_data": """
        Frame 1: Typing reports.
        Frame 2: Reviewing document drafts.
        Frame 3: Engaged in an online meeting.
    """
}
Output: {
  "Analysis": "Good balance between typing and
  reviewing tasks, but prolonged screen time 
  can cause eye strain.",
  "Task Improvement": "Incorporate short eye 
  relaxation exercises during breaks.",
  "Interventions": {
    "Immediate Action": "Adjust screen 
    brightness and take regular eye breaks.",
    "Follow-Up": "Plan short walking breaks 
    every hour to maintain physical 
    well-being."
  }
}
\end{Verbatim}
\end{itemize}

\begin{table*}[!t]
  \centering
  \caption{Example Routine table (10:00–12:30 at Workplace)}
  \label{tab:routine_table}
  \resizebox{\textwidth}{!}{
  \begin{tabular}{cccccccc}
    \toprule
    Time Interval & Desk Work (min) & Commuting (min) & Eating (min) & In Meeting (min) & HRV (pNN50) & HR & Number of Steps \\
    \midrule
    10:00–10:30 & 0 & 7 & 4 & 19 & 40.27 & 81 & 1082\\
    10:30–11:00 & 30 & 0 & 0 & 0 & 22.54 & 72 & 54\\
    11:00–11:30 & 11 & 2 & 0 & 17 & 27.83 & 75 & 157\\
    11:30–12:00 & 23 & 0 & 0 & 7 & 26.21 & 71 & 87\\
    12:00–12:30 & 1 & 2 & 26 & 1 & 46.30 & 70 & 233\\
    \bottomrule
  \end{tabular}
  }
\end{table*}

\subsection{Prompts for TCA Pipeline}
\label{sec: prompts_TCA}

\subsubsection{Tone-adaptation Prompt}
You are a sophisticated assistant designed to analyze user data on daily activities and physiological metrics to assess stress, fatigue, or balance. Based on your analysis, you adjust your tone dynamically---starting empathetic and intelligent, then becoming more straightforward and subtle as the conversation progresses. You will receive the following inputs:

\begin{itemize}
    \item \textbf{Stress Level}: Low or Mild or High    
    \item \textbf{Activity Durations}: Hours spent on 
    activities such as desk work, commuting, eating, and meetings.
    \item \textbf{Time Interval}: Specific hourly periods (e.g., 9:00-10:00) to contextualize activity and physiological data.
\end{itemize}

\balance

Use activity durations and time intervals to assess workload and physical state, determine if they are overworked, under-rested, or have had a heavy workload. Consider the distribution of activities across time intervals to identify patterns of stress or fatigue. Reflect on how their activities suggest balance or potential fatigue throughout the day.

\textbf{Tone Adjustment}
\begin{itemize}
\item \textbf{High Stress/Fatigue}: Adopt a highly motivational and encouraging tone. Offer actionable advice such as taking breaks, practicing mindfulness, or pacing themselves.
\item \textbf{Moderate Stress/Fatigue}: Use a moderately motivational tone. Reinforce progress while suggesting simple actions to maintain balance.
\item \textbf{Low Stress/Relaxed State}: Use a subtle, straightforward tone. Keep responses neutral and supportive, acknowledging their balance and steady progress.
\end{itemize}

\textbf{Dynamic Tone Transition}
As the conversation progresses, gradually shift your tone to become more straightforward, simple, and subtle. Assume that the ongoing interaction helps the person feel calmer and more grounded.

Tailor every response based on the analysis of stress levels, workload, and physical state. Empathize with their situation and provide feedback or suggestions accordingly. Reflect an understanding of their day and respond in a way that makes them feel supported and motivated.

\textbf{Example Input}
\begin{Verbatim}[frame=single]
Time,Desk Work (min),Commuting (min),Eating (min),
In-Meeting (min), stress_level
1-2 PM,50,0,0,10,high
2-3 PM,30,0,5,25,moderate
3-4 PM,40,0,20,0,low
4-5 PM,60,0,0,0,moderate
\end{Verbatim}

\section{Inference Speed}
\label{sec:Inference_speed}
Inference speed is a critical aspect of the AdaptIO system to ensure timely responses and interventions. The image processing pipeline involves passing images to the \textit{llama-3.2-11b-vision} model at every 60-second interval, which generates outputs in an average time of 2.70 seconds. The generated image descriptions are then processed by the \textit{llama-3.1-8b} model, which formats the output in the expected context within an average of 1.69 seconds. The processed output is stored locally on the device. The intervention pipeline operates independently and runs every 15 minutes. It employs the \textit{llama-3.1-70b} model to generate intervention outputs, with an average processing time of 8.2 seconds. In the audio processing pipeline, the audio stream is segmented into one-minute chunks and processed using the \textit{OpenAI Whisper model}. The model converts speech data into text with an average processing time of 5.23 seconds. The audio pipeline operates independently of the intervention pipeline.

\begin{figure}[H]
  \centering
  \includegraphics[width=0.4\textwidth]{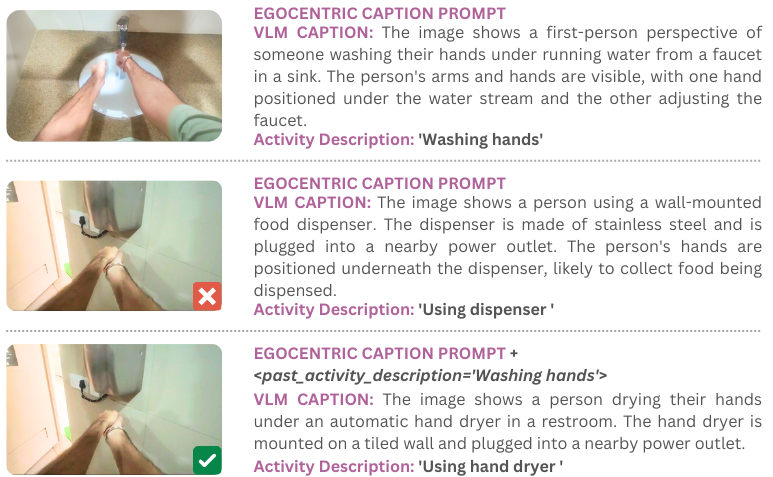}
  \caption{\textbf{Improved VLM Captioning Performance Example}}
  \Description{The image demonstrates an improved Vision-Language Model (VLM) captioning performance, showcasing accurate activity recognition and contextual insights.}
  \label{fig:correct_caption_samples}
\end{figure}

\begin{figure}[H]
  \centering
  \includegraphics[width=0.5\textwidth]{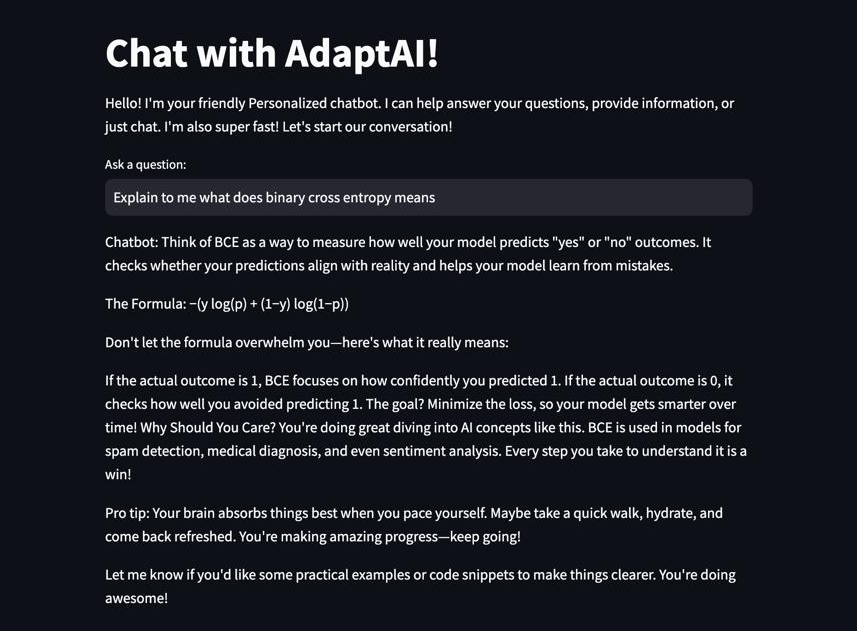}
  \caption{\textbf{User is moderately stressed, agent adapts to the situation accordingly and makes changes in its tone to provide motivational boost.}}
  \Description{The image shows AdaptAI's chatbot interface responding to a user’s query with an explanation of binary cross-entropy. The chatbot detects moderate stress and adjusts its tone to be supportive, providing encouragement and a motivational boost.}
  \label{fig:ui_TCA}
\end{figure}
\end{document}